\documentclass[preprint,showpacs,showkeys,preprintnumbers,amsmath,amssymb]{revtex4-2}

\usepackage{graphicx}
\usepackage{dcolumn}
\usepackage{bm}

\begin{document}

\title{Fast convergence to an approximate solution by  
message-passing for complex optimizations
}

\author{Yukio Hayashi}
\affiliation{Division of Transdisciplinary Sciences,
Graduate School of Advanced Science and Technology, 
Japan Advanced Institute of Science and Technology,\\
Ishikawa, 923-1292, Japan}

\date{\today}

\begin{abstract}
  Message-passing (MP) is a powerful tool for finding an
  approximate solution in optimization.
  We generalize it to nonlinear product-sum form, and numerically
  show the fast convergence for the minimum feedback vertex set
  and the minimum vertex cover known as NP-hard problems.
  From the linearity of MP in a logarithmic space,
  it is derived that an equilibrium solution exists in
  a neighborhood of random initial values.
  These results will give one of the reason why
  the convergence is very fast in collective computation
  based on a common mathematical background.
\end{abstract}

%



\maketitle


\newpage
\section{Introduction}
In complex optimization,
one of the important issue is investigating
how fast an approximate solution can be obtained.
For example, 
by learning of connection weight parameters on a neural network,
it is the task to search a target function
from input to output in a high-dimensional pattern space
in order to minimize 
the square error between output and teacher signals.
In general,
there are many local minima obtained by 
learning of neural network.
Surprisingly, it has been shown with much interest
that any target function can be realized in a sufficiently
small neighborhood of random weight parameters on a neural network
with sufficiently wide hidden layers through learning based
on signal propagation \cite{Jacot18,Lee19,Amari20a,Amari20b}. 
Instead of the complicated analysis \cite{Jacot18,Lee19},
as an intuitive explanation,
the elementary proof has been presented by applying a linear
theory \cite{Amari20a,Amari20b}.
In other words, for the mapping function of input and output patterns, 
any solution exists in the neighborhood of random weights on a
neural network,
therefore the fast convergence is expected by learning of
random neural network
without wandering in the seach space forever.

On the other hand, from statistical physics approach,
similar but different propagation methods called
message-passing (MP) have been developed \cite{Mezard09}.
Although there are some types of MP 
with a same name: {\it belief propagation} (BP),
they are not equivalent.
One is well-known BP \cite{Yedidia01}
to decode low-density parity-check codes \cite{Gallager63}
or to restore a damaged image 
on a graphical model \cite{Weiss00,Mezard09}.
This type of MP
is represented by sum-product or max-sum form \cite{Shah13}.
Another is to approximately solve 
combinatorial optimization problems \cite{Weigt06,Zhou13}.
That type of MP
is represented by product-sum form as mentioned later.
Through iterations of MP for minimizing a free-energy, 
the former performs propagation of a belief of state,
while the later performs interactions among states.
These different objectives may appear as 
sum-product or max-sum and product-sum forms, respectively.
Here, a state means $\pm 1$ spin in Ising model
\cite{Mezard01,Mezard09}, or 
nodes' label such as root identifier excluded 
in feedback vertex set (FVS) \cite{Zhou13}
or covered/uncovered node's state as the candidate
included/excluded in a set of vertex cover (VC) \cite{Weigt06}.
FVS is a set of nodes that are necessary to form loops, 
while VC is a set of covered nodes that are at least one of end-nodes
for each link.
Their approximate solutions are applicable to extract
influencers \cite{Liao22} and to enhance robustness of connectivity
\cite{Chujyo21} in complex networks.
Theoretically 
an unique solution is assured by MP 
on only a tree, however practically 
a good solution is obtained by MP
on a network with (long) loops in many cases.
In general, on a loopy network, 
there can be many equilibrium solutions or
sometimes more complex oscillations by MP
depending on initial values.

In this paper,
not only the fast convergence by MP in product-sum form 
is numerically shown,
but also it is derived as a reason of fast convergence
that
the solution exists in a neighborhood of random initial values,
based on a common mathematical background of linear theory
\cite{Amari20a}.
We take particular note of that
approximately solving optimization problems by MP and
learning of neural network are related as propagation methods
for collective computation,
in spite of studying in different research fields of
information science or machine learning and statistical physics
or network science.

\section{Message-passing in product-sum form}
We briefly review the statistical physics approaches 
\cite{Zhou13,Weigt06}
to find approximate solutions for some of 
combinatorial optimizations such as the minimum
FVS and VC problems.
However, to seek common ground,
the representations are slightly modified for generalizing
them to MP in product-sum form.
Note that the left-hand side in MP equation 
is updated by substituting the right-hand side, repeatedly.

\subsection{For the minimum feedback vertex set}
By using a cavity method \cite{Mezard01,Mezard09} for 
estimating the minimum FVS known as NP-hard \cite{Karp72}, 
it is assumed that 
nodes $v \in \partial u$ are mutually independent of each other,
when node $u$ is removed. 
Here, $\partial u$ denotes the set of connecting neighbor nodes 
of $u$.
In the cavity graph, 
if all nodes $v \in \partial u$ are either empty ($A_{v} = 0$)
or roots ($A_{v} = v$), 
the added node $u$ can be a root ($A_{u} = u$).
There are the following exclusive states \cite{Zhou13}.
\begin{description}
  \item[State $A_{u} = 0$:] $u$ is empty. 
    Since $u$ is unnecessary as a root, it belongs to FVS.
  \item[State $A_{u} = u$:] $u$ becomes its own root.\\
    The state $A_{v} = v$ of $v \in \partial u$ 
    is changeable to $A_{v} = u$, when node $u$ is added.
  \item[State $A_{u} = w$:] one node $w \in \partial u$ becomes the root
    of $u$, when it is added, 
    if $w$ is occupied and all other $w' \in \partial u \backslash w$ 
    are either empty or roots.
\end{description}

For a link $u \rightarrow v$, $v \in \partial u$, 
the corresponding probabilities to the above states are represented
by the following MP equations \cite{Zhou13}.
\begin{equation}
  q^{0}_{u \rightarrow v} = \frac{e^{-x}}{z^{FVS}_{u \rightarrow v}(t)}
  = \frac{\Pi_{w \in \partial u \backslash v}
    \sqrt[d_{u}-1]{e^{-x}}}{z^{FVS}_{u \rightarrow v}(t)},
  \label{eq_update_quv^0}
\end{equation}
\begin{equation}
   q^{u}_{u \rightarrow v} = 
  \frac{\Pi_{w' \in \partial u \backslash v} 
    \left( q^{0}_{w' \rightarrow u} + q^{w'}_{w' \rightarrow u} \right)
  }{z^{FVS}_{u \rightarrow v}},  \label{eq_update_quv^u}
\end{equation}
\begin{equation}
  q^{w}_{u \rightarrow v} = \frac{(1 - q_{w \rightarrow u}^{0})
    \Pi_{w' \in \partial u \backslash v, w} (q_{w' \rightarrow u}^{0}
    + q_{w' \rightarrow u}^{w'}) }{z^{FVS}_{u \rightarrow v}},
  \;\;\; w \in \partial u \backslash v, 
  \label{eq_update_quv^w}
\end{equation}
where $d_{u}$ denotes the degree of node $u$,
$\partial u \backslash v$ is the subset of $\partial u$ except $v$,
and $x > 0$ is a parameter of inverse temperature
to give a penalty $e^{-x}$ for minimizing the size of FVS.
We have the normalization constant 
\begin{equation}
  z^{FVS}_{u \rightarrow v} \stackrel{\rm def}{=} 
  e^{-x} + \left\{ \Pi_{w' \in \partial u \backslash v} 
  \left( q^{0}_{w' \rightarrow u} + q^{w'}_{w' \rightarrow u} \right)
  \times \left( 1 + \sum_{w \in \partial u \backslash v} 
    \frac{1 - q^{0}_{w \rightarrow u}}{
      q^{0}_{w \rightarrow u} + q^{w}_{w \rightarrow u}} \right) \right\}, 
  \label{eq_z_uv_FVS}
\end{equation}
to satisfy
\begin{equation}
q^{0}_{u \rightarrow v} + q^{u}_{u \rightarrow v} 
  + \sum_{w \in \partial u \backslash v} q^{w}_{u \rightarrow v} = 1.
\label{normalize_FVS}
\end{equation}
Note that there are $d_{u}-1$ links of $w' \rightarrow u$
except $v \rightarrow u$, 
and that the multiplication term $\times 1$ 
of Eq.(\ref{normalize_FVS}) is hidden in the numerator 
of the right-hand side of Eq.(\ref{eq_update_quv^0}).
$1 - q^{0}_{w \rightarrow u} = q^{1}_{w \rightarrow u}
+ \sum_{w' \in \partial w \backslash u} q^{*w'}_{w \rightarrow u}$
is also a sum form.
Thus, the right-hand sides of Eqs.
(\ref{eq_update_quv^0})(\ref{eq_update_quv^u})(\ref{eq_update_quv^w})
are product-sum forms.

The $0$ state probability of $u$ included in FVS
is given by \cite{Zhou13}
\begin{equation}
q_{u}^{0} \stackrel{\rm def}{=}
\frac{e^{-x}}{e^{-x} + \left\{ 1 + \sum_{w \in \partial i}
  \frac{1 - q_{w \rightarrow u}^{0}}{q_{w \rightarrow u}^{0}
    + q_{w \rightarrow u}^{w}} \right\} \Pi_{v \in \partial i}
  \left( q_{v \rightarrow u}^{0} + q_{v \rightarrow u}^{v} \right) }.
\label{eq_def_qu^0}
\end{equation}

\subsection{For the minimum vettex cover}
In the cavity graph for estimating the minimum VC
known as NP-hard \cite{Karp72},
since at least one end-node of each link should be covered,
the following three exclusive states at node $u$
are considered 
for a link $u \rightarrow v$, $v \in \partial u$ 
\cite{Weigt06}.
\begin{description}
\item[Sate $0$:] $u$ is uncoverd, when there are no uncovered nodes
  $w \in \partial u \backslash v$.
\item[Sate $1$:] $u$ is coverd, when two or more nodes
  $w \in \partial u \backslash v$ are uncovered.
\item[Sate $*w$:] As joker state, $u$ is sometimes covered and
  sometimes uncovered, when only one node
  $w \in \partial u \backslash v$ is uncovered but other
  $w' \in \partial u \backslash v, w$ are covered or joker states.
\end{description}

The corresponding probabilities are represented
by the following MP equations \cite{Weigt06}.
\begin{equation}
  q^{0}_{u \rightarrow v} =
  \frac{\Pi_{w' \in \partial u \backslash v}
  (1 - q^{0}_{w' \rightarrow u})}{z^{VC}_{u \rightarrow v}}, 
  \label{eq_update_puv^0}
\end{equation}
\begin{eqnarray}
  q^{1}_{u \rightarrow v} & = & \frac{e^{-x}
  \left[ 1 - \Pi_{w' \in \partial u \backslash v} (1 - q^{0}_{w' \rightarrow u})
    - \sum_{w \in \partial u \backslash v} q^{0}_{w \rightarrow u}
    \Pi_{w' \in \partial u \backslash v,w} (1 - q^{0}_{w' \rightarrow u})
    \right]}{z^{VC}_{u \rightarrow v}}, \nonumber \\
  & = & \frac{\Pi_{w' \in \partial u \backslash v} \sqrt[d_{u}-1]{
      e^{-x} \mbox{(the numerator of above equation)}
  }}{z^{VC}_{u \rightarrow v}}, \label{eq_update_puv^1}
\end{eqnarray}
\begin{eqnarray}
  q^{*w}_{u \rightarrow v} & = &
  \frac{e^{-x} q^{0}_{w \rightarrow u}
    \Pi_{w' \in \partial u \backslash v,w} (1 - q^{0}_{w' \rightarrow u})}{
    z^{VC}_{u \rightarrow v}}, \nonumber \\
  & = & \frac{\sqrt[d_{u}-1]{e^{-x}} q^{0}_{w \rightarrow u}
  \Pi_{w' \in \partial u \backslash v,w} \left[ \sqrt[d_{u}-1]{e^{-x}}
    (1 - q^{0}_{w' \rightarrow u}) \right]}{z^{VC}_{u \rightarrow v}},
  \;\;\; w \in \partial u \backslash v, 
  \label{eq_update_puv^*}
\end{eqnarray}
with the normalization constant 
\begin{equation}
  z^{VC}_{u \rightarrow v} \stackrel{\rm def}{=} e^{-x} \left[ 1 - (1 - e^{x})
    \Pi_{w' \in \partial u \backslash v} (1 - q^{0}_{w' \rightarrow u}) \right],
  \label{eq_z_uv_VC}
\end{equation}
to satisfy
$q^{0}_{u \rightarrow v} + q^{1}_{u \rightarrow v} +
\sum_{w \in \partial u \backslash v} q^{*w}_{u \rightarrow v} = 1$.
From this normalization condition and
$1 - q_{u \rightarrow v}^{0} = q^{1}_{u \rightarrow v} +
\sum_{w \in \partial u \backslash v} q^{*w}_{u \rightarrow v}$,
the right-hand sides of 
Eqs.(\ref{eq_update_puv^0})(\ref{eq_update_puv^*})
are product-sum forms.
The right-hand sides of Eqs.(\ref{eq_update_puv^1}) may be not
exactly the form.
However, Eq.(\ref{eq_update_puv^0}) is essential and other ancillary
Eqs.(\ref{eq_update_puv^1})(\ref{eq_update_puv^*}) are not,
since $z^{VC}_{u \rightarrow v}$ in Eq.(\ref{eq_z_uv_VC}) is represented
by only the $0$ state's probabilities.

The $1$ state probability of $u$ included in VC
is given by \cite{Weigt06}
\begin{equation}
q_{u}^{1} \stackrel{\rm def}{=}
\frac{e^{-x} \left\{ 1 - \Pi_{v \in \partial u} (1 - q_{v \rightarrow u}^{0})
  - \sum_{w \in \partial u} q_{w \rightarrow u}^{0}
  \Pi_{w' \in \partial u \backslash w} (1 - q_{w' \rightarrow u}^{0})
  \right\}}{e^{-x} \left\{ 1 - (1-e^{x})
  \Pi_{v \in \partial u} (1 - q_{v \rightarrow u}^{0}) \right\}}.
\label{eq_def_qu^1}
\end{equation}

\subsection{For more general cases}
We consider a set
$\Omega_{u} = \{ \alpha_{u}, \beta_{u}, \gamma_{u}, \ldots, \kappa_{u}, 
\ldots, \omega_{u} \}$
of states in any order.
The number $| \Omega_{u} |$ can be different for each node $u$.
In the case of minumum FVS,
the states are $\alpha_{u} = 0$, $\beta_{u} = u$, and 
$\gamma_{u}, \ldots, \omega_{u} \in \partial u \backslash v$.
In the case of minumum VC,
the states are $\alpha_{u} = 0$,
and $\gamma_{u}, \ldots, \omega_{u}$ are jokers $*$ of 
$w \in \partial u \backslash v$.
$\beta_{u} = 1$ is an exception by
$q_{u \rightarrow v}^{1} = 1 - (q^{0}_{u \rightarrow v} +
\sum_{w \in \partial u \backslash v} q^{*w}_{u \rightarrow v})$.

Then, by an inspiration from Ref. \cite{Xiao11}, 
Eqs.
(\ref{eq_update_quv^0})(\ref{eq_update_quv^u})(\ref{eq_update_quv^w})(\ref{eq_z_uv_FVS})
or Eqs.
(\ref{eq_update_puv^0})(\ref{eq_update_puv^*})(\ref{eq_z_uv_VC})
are generalized as MP in product-sum form.
In the following example,
we set that only one state $\alpha_{u}$ has a penalty term
$e^{-x}$ to minimize its probability.

\begin{eqnarray}
q^{\alpha_{u}}_{u \rightarrow v}(t+1)
& = & \frac{e^{-x}}{z_{u \rightarrow v}(t)}
\Pi_{w' \in \partial u \backslash v} \left( \sum_{\delta \in S_{\alpha_{u}}}
q^{\delta}_{w' \rightarrow u}(t) \right), \nonumber \\
& = & \frac{1}{z_{u \rightarrow v}(t)}
\Pi_{w' \in \partial u \backslash v} \left( \sqrt[d_{u}-1]{e^{-x}}
\sum_{\epsilon \in S_{\alpha_{u}}} q^{\epsilon}_{w' \rightarrow u}(t) \right), 
\label{eq_update_quv^alpha}
\end{eqnarray}
\[
\hspace{2cm} \vdots
\]
\begin{equation}
q^{\kappa_{u}}_{u \rightarrow v}(t+1) = \frac{1}{z_{u \rightarrow v}(t)}
\Pi_{w' \in \partial u \backslash v} \left( \sum_{\delta' \in S_{\kappa_{u}}}
q^{\delta'}_{w' \rightarrow u}(t) \right),
\label{eq_update_quv^kappa}
\end{equation}
\[
\hspace{2cm} \vdots
\]
\begin{equation}
q^{\omega_{u}}_{u \rightarrow v}(t+1) = \frac{1}{z_{u \rightarrow v}(t)}
\Pi_{w' \in \partial u \backslash v} \left( \sum_{\delta'' \in S_{\omega_{u}}}
q^{\delta''}_{w' \rightarrow u}(t) \right),
\label{eq_update_quv^omega}
\end{equation}
where
$S_{\alpha_{u}}, \ldots, S_{\kappa_{u}},
\ldots, S_{\omega_{u}} \subset \Omega_{u}$,
$w' \in \partial u \backslash v$, and 
$z_{u \rightarrow v}(t)$ is defined to satisfy the normalization condition 
$\sum_{\kappa_{u} \in \Omega_{u}} q^{\kappa_{u}}_{u \rightarrow v}(t) = 1$
at each time $t$.
We emphasize the dependence of iteration time step $t \geq 0$
in the representation of Eqs.
(\ref{eq_update_quv^alpha})(\ref{eq_update_quv^kappa})(\ref{eq_update_quv^omega}).

\section{Numerical and theoretical analysises}
\subsection{Simulation results for fast convergence of MP}
We numerically investigate the convergence of MP Eqs.
(\ref{eq_update_quv^0})(\ref{eq_update_quv^u})(\ref{eq_update_quv^w}) or
(\ref{eq_update_puv^0})(\ref{eq_update_puv^1})(\ref{eq_update_puv^*})
for the minimum FVS or VC problem.
It is evaluated by the cosine similarity $Sim(t)$ 
between the state probabilities at iteration times $t-1$ and $t$,
\[
Sim(t) \stackrel{\rm def}{=}
\frac{\sum_{u=1}^{N} \sum_{e:u \rightarrow v} \sum_{\kappa_{u} \in \Omega_{u}}
  q_{e}^{\kappa_{u}}(t-1) \times q_{e}^{\kappa_{u}}(t)}{\sqrt{
    \sum_{u=1}^{N} \sum_{e:u \rightarrow v} \sum_{\kappa_{u} \in \Omega_{u}}
    q_{e}^{\kappa_{u}}(t-1)^{2} } \times \sqrt{
    \sum_{u=1}^{N} \sum_{e:u \rightarrow v} \sum_{\kappa_{u} \in \Omega_{u}}
    q_{e}^{\kappa_{u}}(t)^{2}} }.
\]
When $Sim(T)$ approaches to $1$,
the state probabilities  $\{ q_{e}^{\kappa_{u}}(T) \}$
converge around a time $T$.

\begin{figure}[ht!]
  \includegraphics[width=.9\textwidth]{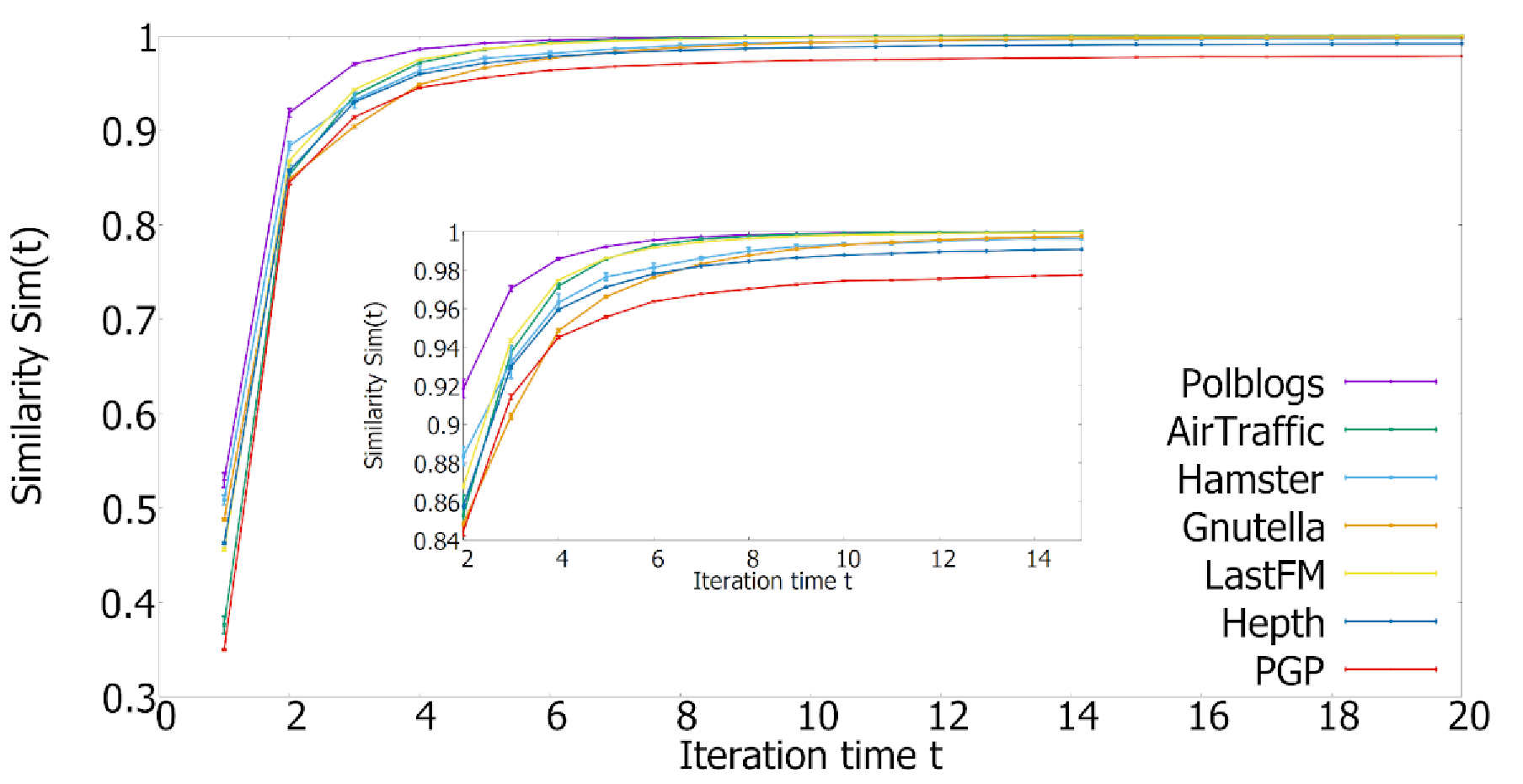}
  \caption{Similarity between the state probabilities 
    at iteration times $t-1$ and $t$ by simple MP for the minimum FVS
    in real networks distinguished by color lines.
  Inset show the enlarged part to see the convergent curves.}
  \label{fig_simTS_FVS}
\end{figure}

\begin{figure}[ht!]
  \includegraphics[width=.9\textwidth]{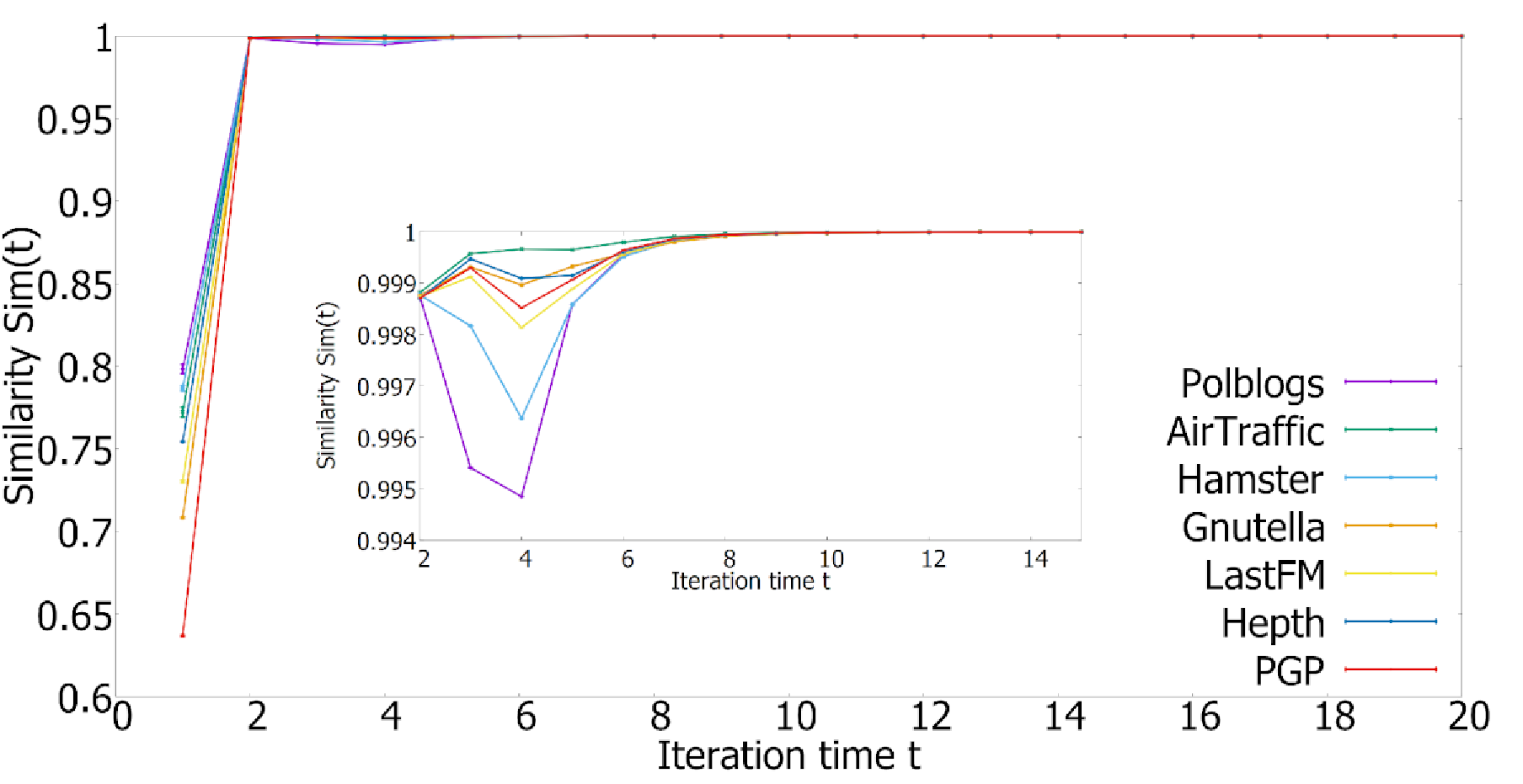}
  \caption{Similarity between the state probabilities 
    at iteration times $t-1$ and $t$ by simple MP for the minimum VC
    in real networks distinguished by color lines.
  Inset show the enlarged part to see the convergent curves.}
  \label{fig_simTS_VC}
\end{figure}

\begin{table}[ht!]
  \caption{Data source.
    Name of network,
    numbers $N$, $M \stackrel{\rm def}{=} \sum_{u=1}^{N} d_{u} / 2$
    of nodes and links, diameter $D$ as the maximum shortest path length
    between two nodes, access destination.
    Polblogs $\sim$ PGP are Scale-Free networks whose degree distributions
    follow power-law, while Cost2666 and Janos-us-ca are planar communication
    networks.
  }
  \label{table_data_source}
  \begin{tabular}{c|ccc|l} 
    Name & $N$ & $M$ & $D$ & URL \\ \hline
    Polblogs & 1222 & 16714 & 8 & http://www-personal.umich.edu/$\sim$mejn/netdata/ \\
    AirTraffic & 1226 & 2408 & 17 & https://data.europa.eu/data/datasets \\ 
    &      &   &  & /12ec37d3-ada7-4d4c-84ef-f347b1d8dedf?locale=fi \\
    Hamster & 1788 & 12476 & 14 & https://networkrepository.com/soc-hamsterster.php \\
    Gnutella & 6299 & 20776 & 9 & http://snap.stanford.edu/data/p2p-Gnutella08.html \\    
    LastFM & 7624 & 278060 & 15 & http://snap.stanford.edu/data/feather-lastfm-social.html \\ 
    Hepth & 8638 & 24806 & 18 & http://snap.stanford.edu/data/ca-HepTh.html \\
    PGP & 10680 & 24316 & 24 & https://deim.urv.cat/$\sim$alexandre.arenas/data/welcome.htm \\ \hline
    Cost266     & 37 & 56 & 8 & http://sndlib.zib.de/download/sndlib-networks-native.zip \\
    Janos-us-ca & 39 & 61 & 10 & http://sndlib.zib.de/download/sndlib-networks-native.zip \\ \hline
  \end{tabular}
\end{table}

\begin{table}
  \center
  \caption{Cosine similarities between
    $\{ q_{e}^{\kappa_{u}}(T^{*}) \}$ and $\{ \tilde{q}_{e}^{\kappa_{u}}(0) \}$ (1st column), and between
    $\{ q_{e}^{\kappa_{u}}(T^{*}) \}$ and $\{ \tilde{q}_{e}^{\kappa_{u}}(T^{*}) \}$ (2nd column).
  The results are averaged over 100 samples.}
  \label{table_perturb}
  \begin{center} (a) FVS \end{center}
  \begin{tabular}{c|ccccccc} \hline
    & Polblog & Airtraffic & Hamster & GNUtella & LastFM & Hepth & PGP \\ \hline
    $\{ q_{e}^{\kappa_{u}}(T^{*}) \}$ and $\{ \tilde{q}_{e}^{\kappa_{u}}(0) \}$
    & 0.768 & 0.883 & 0.874 & 0.936 & 0.934 & 0.949 & 0.877 \\
    $\{ q_{e}^{\kappa_{u}}(T^{*}) \}$ and $\{ \tilde{q}_{e}^{\kappa_{u}}(T^{*}) \}$
    & 0.998 & 0.993 & 0.998 & 0.998 & 0.995 & 0.997 & 0.996 \\ \hline
  \end{tabular}
  \begin{center} (b) VC \end{center}
  \begin{tabular}{c|ccccccc} \hline
    & Polblog & Airtraffic & Hamster & GNUtella & LastFM & Hepth & PGP \\ \hline
    $\{ q_{e}^{\kappa_{u}}(T^{*}) \}$ and $\{ \tilde{q}_{e}^{\kappa_{u}}(0) \}$
    & 0.770 & 0.879 & 0.864 & 0.944 & 0.940 & 0.944 & 0.960 \\
    $\{ q_{e}^{\kappa_{u}}(T^{*}) \}$ and $\{ \tilde{q}_{e}^{\kappa_{u}}(T^{*}) \}$
    & 0.988 & 0.991 & 0.998 & 0.995 & 0.996 & 0.997 & 0.994 \\ \hline
  \end{tabular}
\end{table}

The following results are averaged over 100 samples from initial 
$\{ q_{e}^{\kappa_{u}}(0) \}$
of uniform random numbers in the interval $(0, 1)$.
We set the parameter of inverse temperature as $x=7$.
Note that the unit time consists of the updating by MP 
in order of random permutations of $N$ nodes and $d_{u}$ links 
to avoid vibration behaviour as few as possible
instead of synchronously simultaneous updating of all.

Figures \ref{fig_simTS_FVS} and \ref{fig_simTS_VC}
show the time evolutions of $Sim(t)$ by MP
for the minimum FVS and VC problems, respectively, 
on real networks with thousands nodes and links 
as shown in Table \ref{table_data_source}.
Each colored curves are quickly converged until only several 
iterations less than around ten.
Moreover, the variance indicated by vertical line is very
small in 100 samples.
Such small variance on each colored line means that
similar $Sim(t)$ is obtained at each time $t$ from any initial value,
because the time-course can reach an equilibrium solution in the neighborhood
of random initial value as discussed in the next section.
In other words, without almost depending on the topological difference,
$Sim(t)$ behaves similarly even for the convergence to different
equilibrium solutions which depend on initial values.
On the other hand,
there are different shapes of curves for FVS and VC
in Figs. \ref{fig_simTS_FVS} and \ref{fig_simTS_VC}.
Depending on data in Table \ref{table_data_source},
these colored curves are also slightly different in each of
Figs. \ref{fig_simTS_FVS} and \ref{fig_simTS_VC}.
The tested networks are Scale-Free (SF) commonly but with different 
total numbers $N$, $M$ of nodes and links, and the diameter $D$.
Other topological properties may be different,
however not only huge candidates of topological measures can be considered 
such as clustering coefficient, average length of the shortest paths, 
degree-degree correlations,
modularity or motifs, and so on,
but also it is unestimable which are determinant in advance.
The reasons of different shapes of curves
are considered from the differences of sum terms
or of number of states with penalty in Eqs.
(\ref{eq_update_quv^0})(\ref{eq_update_quv^u})(\ref{eq_update_quv^w})
and 
(\ref{eq_update_puv^0})(\ref{eq_update_puv^1})(\ref{eq_update_puv^*})
and of some topological properties,
although the detail mechanism are unknown at the current stage.
Note that these number of products are same as $d_{u} - 1$ links at node $u$.

In addition,
the existing of equilibrium solution is investigated by random
perturbation for Figs. \ref{fig_simTS_FVS} and \ref{fig_simTS_VC}.
After obtaining an convergent $\{ q_{e}^{\kappa_{u}}(T^{*}) \}$ from any 
$\{ q_{e}^{\kappa_{u}}(0) \}$ of uniform random numbers in the interval $(0, 1)$,
another $\{ \tilde{q}_{e}^{\kappa_{u}}(0) \}$ is set by adding
uniform random numbers in the interval $(-\varepsilon, \varepsilon)$ 
to $\{ q_{e}^{\kappa_{u}}(T^{*}) \}$.
From $\{ \tilde{q}_{e}^{\kappa_{u}}(0) \}$, the corresponding convergent
$\{ \tilde{q}_{e}^{\kappa_{u}}(T^{*}) \}$ is recalculated.
Then, we compute the cosine similarities between
$\{ q_{e}^{\kappa_{u}}(T^{*}) \}$ and $\{ \tilde{q}_{e}^{\kappa_{u}}(0) \}$, and between
$\{ q_{e}^{\kappa_{u}}(T^{*}) \}$ and $\{ \tilde{q}_{e}^{\kappa_{u}}(T^{*}) \}$.
The increased similarities
from first to second columns in Table \ref{table_perturb}(a)(b)
exhibit that, as equilibrium solutions,
same convergent values are almost reached from the neighborhood of them.
Here, we set a sufficient large iteration time $T^{*} = 100$ and
a small perturbation parameter $\varepsilon = 0.4$.
Note that we have also similar results of slightly larger similarities for
$\varepsilon = 0.2$ as closer $\{ \tilde{q}_{e}^{\kappa_{u}}(0) \}$
to $\{ q_{e}^{\kappa_{u}}(T^{*}) \}$.
For Figs. \ref{fig_simTS_FVS} and \ref{fig_simTS_VC},
it is intractable to more regorously analyze the stabilities 
even under a special perturbation of Gaussian distribution, 
because the sizes of Jacobian matrix \cite{Weigt06}
are too large in the linear approximations of Eqs.
(\ref{eq_update_quv^0})(\ref{eq_update_quv^u})(\ref{eq_update_quv^w}) and 
(\ref{eq_update_puv^0})(\ref{eq_update_puv^1})(\ref{eq_update_puv^*})
as nonlinear mappings around equilibrium solutions whose number is unknown.
In the case of FVS, some extensions are required involving complex calculations with
not only $0$ states in the case of VC but also other $u$ and $w$ states,
$w \in \partial u \backslash v$, the analysis will be a further study.
However,
for Cost266 and Janos-us-ca with small sizes in Table \ref{table_data_source},
we calculate it in the case of VC which has only essential variables of $0$ states.
Under the perturbation of Gaussian distribution
in applying the derivation for VC \cite{Weigt06},
we confirm the stabilities of equilibrium solutions
by obtaining that the largest eigenvalues of the following Jacobian matrix $J$ 
are less than $1$ for all $100$ trials of random
initial values after $T^{*} = 10$, $100$, and $1000$ iterations, respectively.
\[
J_{u \rightarrow v, w \rightarrow u } \stackrel{\rm def}{=}
\left\{ \frac{e^{-x}
    \Pi_{w' \in \partial u \backslash v,w} (1 - q_{w' \rightarrow u}^{0}(T^{*}))}{
    (z_{u \rightarrow v}^{VC}(T^{*}))^{2}} \right\}^{2}, \;\;\;
\text{for} \;\; w \in \partial u \backslash v,
\]
\[
J_{u \rightarrow v, w \rightarrow u' } \stackrel{\rm def}{=} 0,\;\;\;
\text{otherwise for} \;\; u' \neq u, \;
u' = u \; \text{and} \; w = v, \;
\text{or} \; u' = u \; \text{and} \; w \notin \partial u.
\]

If an equilibrium solution (or very close solutions)
is obtained by MP even from
different initial values, it may be difficult to distinguish
the nodes included or excluded in FVS or VC by ambiguous values of
$0 < q_{u}^{0}$ or $q_{u}^{1} < 1$.
Thus,
in practical point of view,
decimation process \cite{Zhou13} is usually performed
for finding the candidate of FVS or VC one by one (or
the candidates of some nodes at once for efficiency),
in which
the unit time consists of $T > 1$ rounds of updating
by MP in order of random permutations of $N$ nodes and $d_{u}$ links.
At each time by decimation process,
the selected node $u$ with the highest $q_{u}^{0}$ or $q_{u}^{1}$
defined by Eq.(\ref{eq_def_qu^0}) or (\ref{eq_def_qu^1})
is removed as candidate of FVS or VC.
As the candidates,
we can also select the highest top dozens of nodes at once.
After removing the selected nodes,
$T$ rounds of updating are performed again at next time.
Such process is repeated until satisfying the condition of no loop
or covering one of end-nodes for each link.
The obtained results with decimation are labels $0/1$ of nodes 
included/excluded in FVS or VC,
they may differ form an equilibrium solution with ambiguous values
in $(0, 1)$ by simple MP without decimation.

\begin{figure}[ht!]
  \includegraphics[width=.9\textwidth]{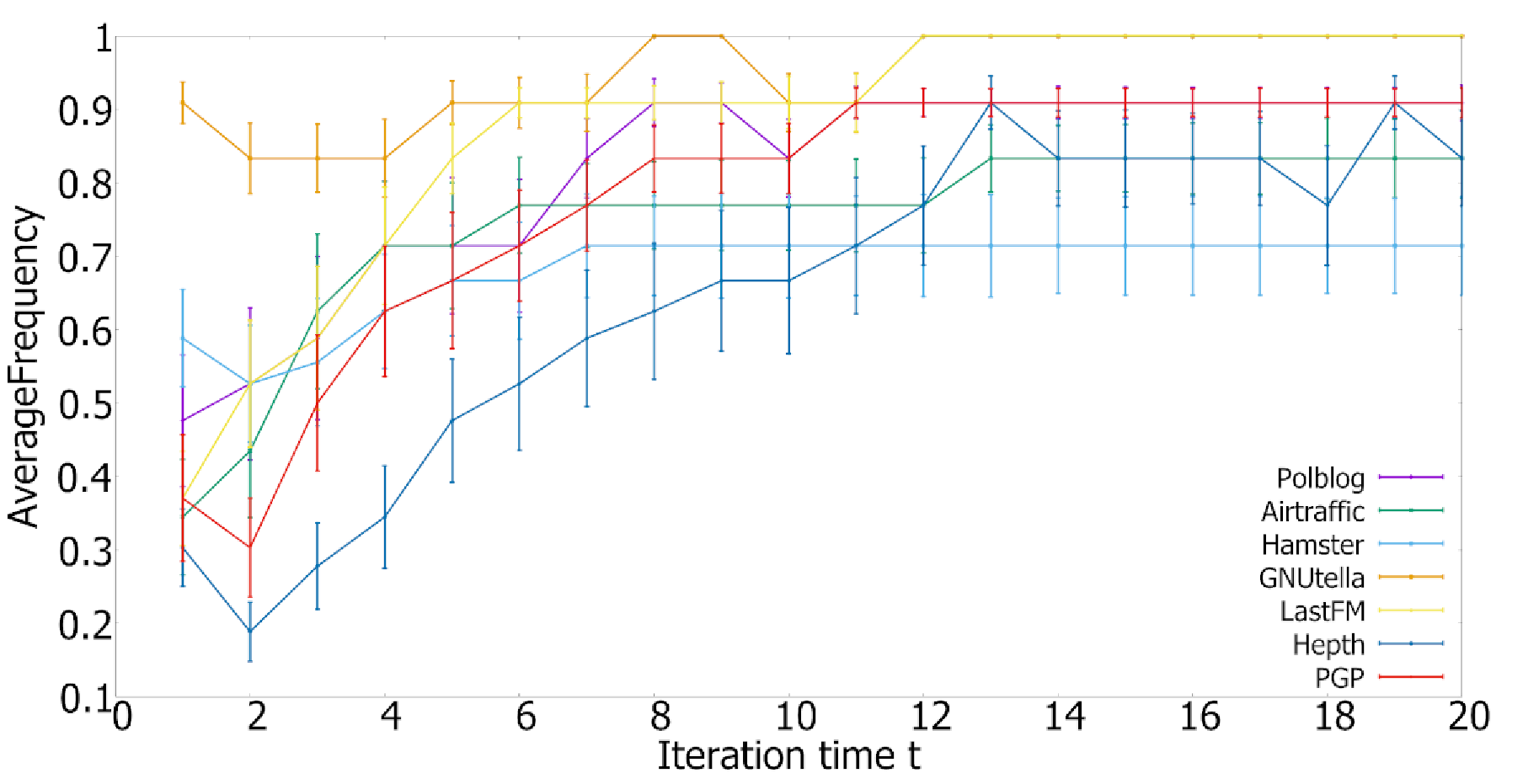}
  \caption{Average frequency of selected nodes as FVS in top 10
    by MP with decimation 
    in real networks distinguished by color lines.
  The vertical bar denotes the variance over 100 samples.}
  \label{fig_AveFreq_FVS}
\end{figure}

\begin{figure}[ht!]
  \includegraphics[width=.9\textwidth]{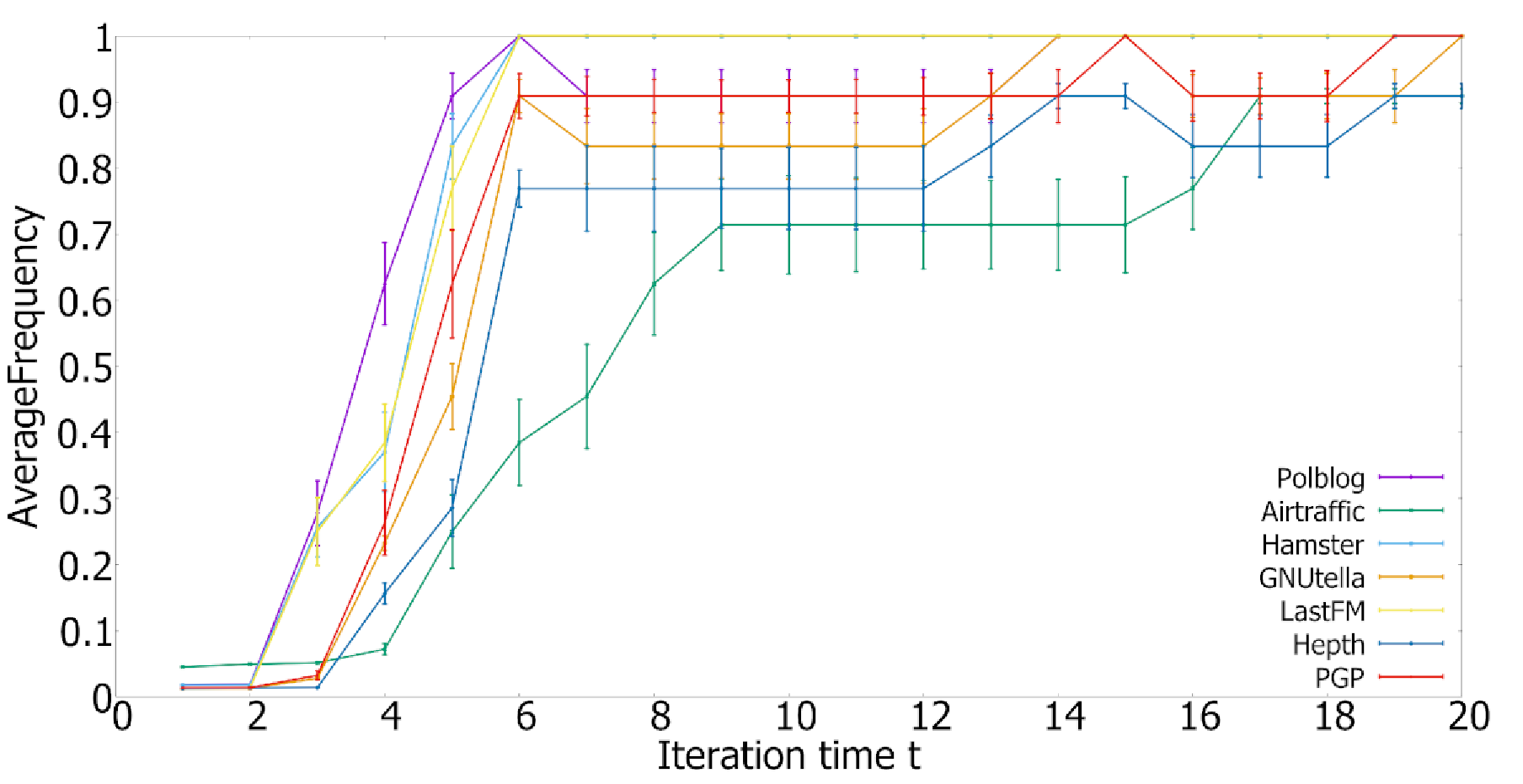}
  \caption{Average frequency of selected nodes as VC in top 10
    by MP with decimation 
    in real networks distinguished by color lines.
  The vertical bar denotes the variance over 100 samples.}
  \label{fig_AveFreq_VC}
\end{figure}

Figures \ref{fig_AveFreq_FVS} and \ref{fig_AveFreq_VC}
show the time evolution of 
average frequency of selected nodes in the highest top 10
at time $t$ by MP with decimation 
over 100 samples of different random initial values.
Commonly, the frequency tends to increase as larger $t$,
however there are slightly different shapes of curves with
different lengths of bars as the variances.
For these differences, 
the reasons are also considered from the differences
of sum terms or of number of states with penalty in Eqs.
(\ref{eq_update_quv^0})(\ref{eq_update_quv^u})(\ref{eq_update_quv^w})
and 
(\ref{eq_update_puv^0})(\ref{eq_update_puv^1})(\ref{eq_update_puv^*})
and of some topological properties.
Here, the maximum frequency $\frac{100}{100}$ means that
all selected nodes are completely overlapped, 
while the minimum frequency $\frac{1}{100}$ means that
selected nodes are non-overlapped and appeared for only one sample.
A value between the maximum and the minimum gives the commonality 
of nodes selected with decimation over samples.
In other words,
it is corresponded to the variety of intermediate stages until
reaching a solution in ranging from unique 
to quite different according to initial values.

\begin{figure}[ht!]
 \begin{minipage}{.496\textwidth}
   \includegraphics[width=\textwidth]{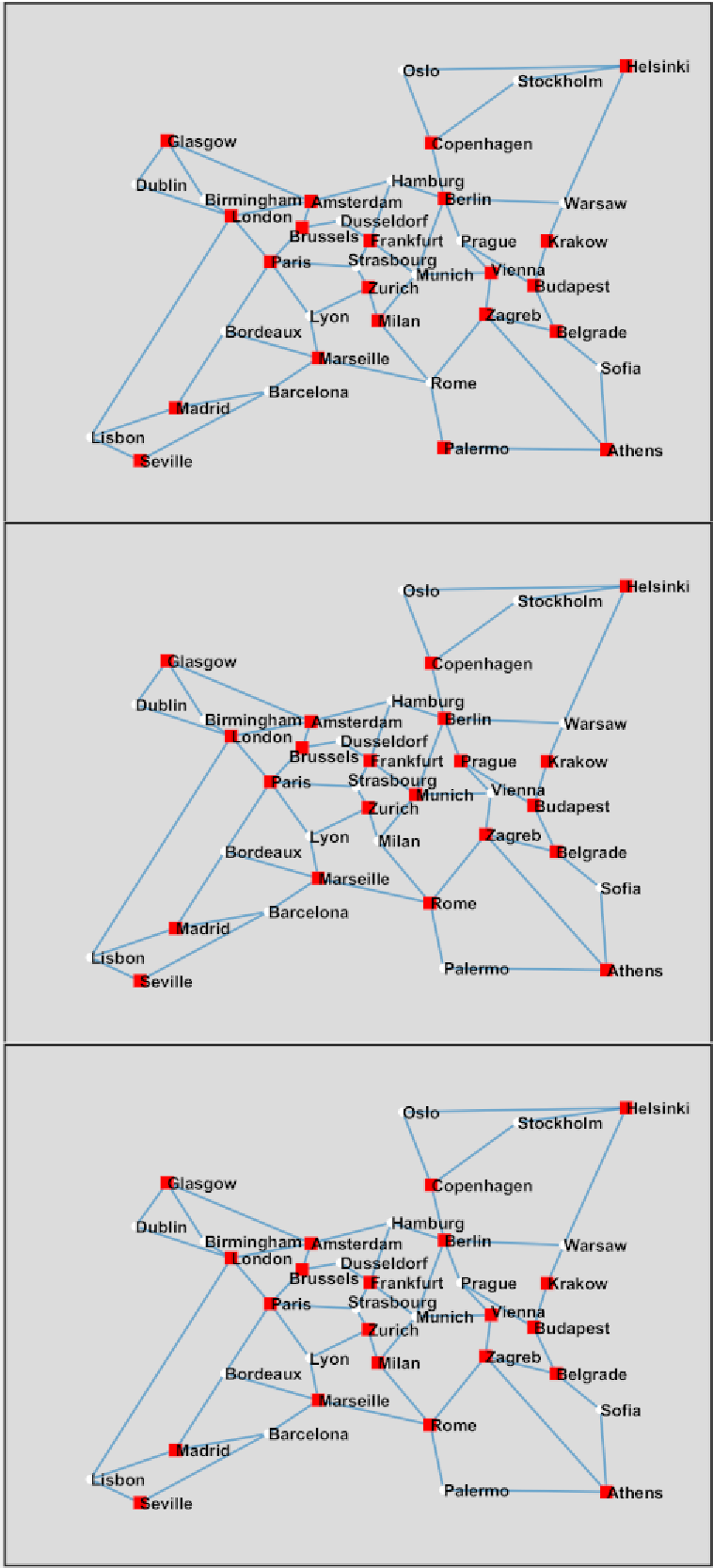}
  \begin{center} (a) Cost266 \end{center}
 \end{minipage}
 \hfill 
 \begin{minipage}{.496\textwidth}
   \includegraphics[width=\textwidth]{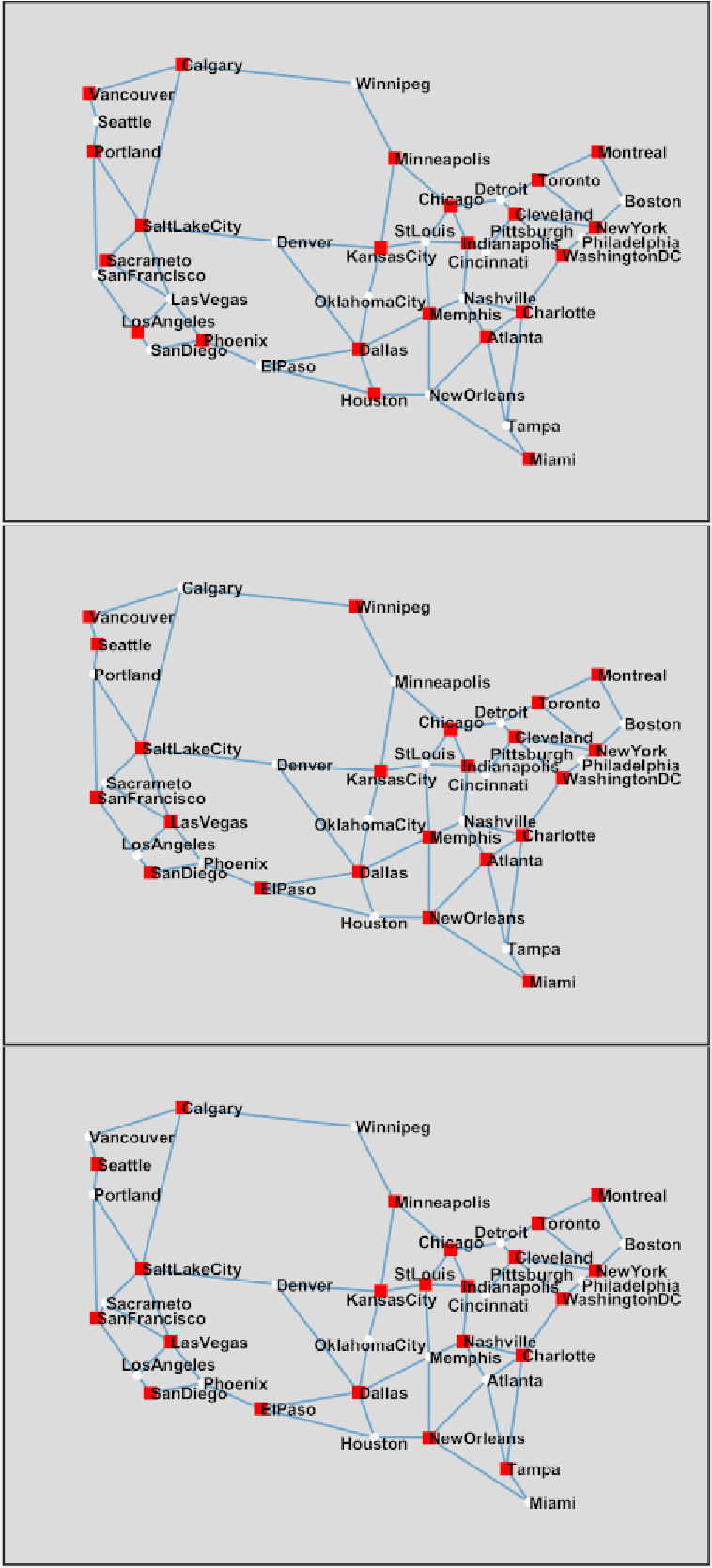}
  \begin{center} (b) Janos-us-ca \end{center}
 \end{minipage}
  \caption{Different sets of VC colored by red from top to bottom 
  according to initial values
  on real communication networks in (a) EU and (b) USA.}
  \label{fig_RealComNet_VC}
\end{figure}

In fact,
as visualized examples in Fig. \ref{fig_RealComNet_VC}
from top to bottom, 
different sets of VC are found by decimation process for $T=10$
on real communication networks (see Table \ref{table_data_source}).
The candidate node is chosen one by one at each time.
Consequently, 
the solutions of VC depend on initial values of state probabilities.
Note that the feasible solution by MP with decimation \cite{Weigt06}
is nearly optimal \cite{Liao22},
since its size is almost half of that 
by a 2-approximation algorithm theoretically guaranteed
in computer science \cite{Bar-Yehuda85}.

\subsection{Linear theory for mesagge-passing in product-sum form}
In this subsection, for MP Eqs.
(\ref{eq_update_quv^alpha})(\ref{eq_update_quv^kappa})(\ref{eq_update_quv^omega})
in product-sum form without round and decimation process,
we study how far is the equilibrium solution from initial values
in the logarithmic space of state probabilities $\{q_{e}^{\kappa_{u}} \}$.
Since Eqs.
(\ref{eq_update_quv^alpha})(\ref{eq_update_quv^kappa})(\ref{eq_update_quv^omega})
are generalizations of Eqs.
(\ref{eq_update_quv^0})(\ref{eq_update_quv^u})(\ref{eq_update_quv^w})
or Eqs.
(\ref{eq_update_puv^0})(\ref{eq_update_puv^*}),
the same discussion is true for 
the case of minimum FVS \cite{Zhou13}
or VC \cite{Weigt06}.
The linear theory is applied by a similar but slightly different
way to learning of multilayer neural networks \cite{Amari20a}
(See Appendix A.1 for the brief review).
In advance, 
we should take care of that
only the existence of solution is discussed
in a neighborhood of random initial values 
without taking into account dynamics of the trajectory to it
as similar to the case of learning of neural networks 
\cite{Amari20a}.

To eliminate the denominator $z_{u \rightarrow v}$ of partition function
in the right-hand side of Eqs.
(\ref{eq_update_quv^alpha})(\ref{eq_update_quv^kappa})(\ref{eq_update_quv^omega}), 
we consider the logarithm of ratio
$q_{e:u \rightarrow v}^{\kappa_{u}}(t+1) / q_{e:u \rightarrow v}^{\omega_{u}}(t+1)$
in the left-hand side.
Then, from the right-hand side, 
\begin{equation}
  b_{\varepsilon}(t) \stackrel{\rm def}{=}
  \sum_{w' \in \partial u \backslash v} \log \left(
  \frac{\sum_{\delta'_{u} \in S_{\kappa_{u}}}
    q_{w' \rightarrow u}^{\delta'_{u}}(t)}{\sum_{\delta''_{u} \in S_{\omega_{u}}}
    q_{w' \rightarrow u}^{\delta''_{u}}(t)}
  \right),
  \label{eq_def_b}
\end{equation}
is obtained as each element of $K$-dimensional vector $\bm{b}(t)$.
The number $\varepsilon$ depends on link $e:u \rightarrow v$
and state $\kappa_{u} \in \Omega_{u} \backslash \omega_{u}$.
There exist $d_{u}$ links emanated from node $u$,
which has $| \Omega_{u} |$ states.
The total number $K$ of variables is
$\sum_{u=1}^{N} d_{u} \times (| \Omega_{u} | - 1)$,
where $-1$ is the reducing due to the 
denominator w.r.t $\omega_{u}$ in each element of $\bm{b}(t)$.
In the case of minimum FVS or VC,
the states are $0$, $u$ or exception $1$, and $w$ or $*w$,
$w \in \partial u \backslash v$,
we have
$| \Omega_{u} | - 1 = 1 + 1 + d_{u} - 1 -1 = d_{u}$ or $d_{u} -1$
and therefore $K = \sum_{u=1}^{N} d_{u}^{2}$
or $\sum_{u=1}^{N} d_{u} (d_{u} - 1)$.

We also consider a $(K+2M)$-dimensional vector 
\begin{eqnarray}
\bm{y(t)} & \stackrel{\rm def}{=} & \left\{
\log q_{e_{1}}^{\alpha_{1}}(t+1), \ldots, \log q_{e_{1}}^{\omega_{1}}(t+1),
\ldots,
\log q_{e}^{\alpha_{u}}(t+1), \ldots, \log q_{e}^{\omega_{u}}(t+1),
\right. \nonumber \\
& & \left.
\ldots,
\log q_{e_{2M}}^{\alpha_{N}}(t+1), \ldots, \log q_{e_{2M}}^{\omega_{N}}(t+1),
\right\}, \label{eq_def_y}
\end{eqnarray}
and a $K \times (K+2M)$ block-diagonal matrix ${\cal Q}$,
whose submatrix is same as
\begin{equation}
Q_{(e)} \stackrel{\rm def}{=}
\left[
\begin{array}{ccccc}
  1      & 0      & \ldots & 0      & -1\\
  0      & 1      & \ldots & 0      & -1\\
  \vdots & \vdots & \ddots & \vdots & \vdots\\
  0      & 0      & \ldots & 1      & -1
\end{array}
\right]
\label{eq_def_Qe}
\end{equation}
except the size $(|\Omega_{u}|-1) \times |\Omega_{u}|$
for link $e:u \rightarrow v$.

Based on the above preparations,
the logarithm of ratio of Eqs.
(\ref{eq_update_quv^alpha}) or (\ref{eq_update_quv^kappa})
and (\ref{eq_update_quv^omega}) becomes
the following systems of linear equations.
\[
\begin{pmatrix}
    \boxed{Q_{(e_{1})}}
                   &       &              &        & \\
                   & \ddots &              &  \text{\huge{0}} & \\
                   &        & \boxed{Q_{(e)}} &        & \\
                   & \text{\huge{0}} &              & \ddots & \\
                   &        &              &        & \boxed{Q_{(e_{2M})}}
\end{pmatrix}
\left(
  \begin{array}{c}
    \log q_{e_{1}}^{\alpha_{1}}(t+1) \\
    \vdots \\
    \log q_{e_{1}}^{\omega_{1}}(t+1) \\
    \vdots \\
    \log q_{e}^{\alpha_{u}}(t+1) \\
    \vdots \\
    \log q_{e}^{\omega_{u}}(t+1) \\
    \vdots \\
    \log q_{e_{2M}}^{\alpha_{N}}(t+1) \\
    \vdots \\
    \log q_{e_{2M}}^{\omega_{N}}(t+1) 
  \end{array}
\right) =
\left(
  \begin{array}{c}
    b_{1}(t) \\
    \vdots \\
    b_{\varepsilon}(t) \\
    \vdots \\
    b_{K}(t) 
  \end{array}
\right).
\]
We assume that directional links are properly ordered as
$e_{1}: 1 \rightarrow v' \neq 1, \ldots,
e: u \rightarrow v, \ldots,
e_{2M}: N \rightarrow v'' \neq N$
with numbering nodes from $1$ to $N$.
Note that the $\pm 1$ elements of ${\cal Q}_{(e)}$ correspond to
$+ \log q_{e}^{\kappa_{u}}(t+1)$ and $- \log q_{e}^{\omega_{u}}(t+1)$
as the state probability
and the constraint of normalization, respectively,
for each link.
As the matrix-vector form, we have 
\begin{equation}
  {\cal Q} \bm{y}(0) = \bm{b}(0),
  \label{eq_Qy0}
\end{equation}
\begin{equation}
  {\cal Q }\bm{y}(\infty) = \bm{b}(\infty),
  \label{eq_Qyinf}
\end{equation}
where 
$\{ q_{u \rightarrow v}^{\kappa_{u}}(\infty) \}$
means an equilibrium solution for MP Eqs.
(\ref{eq_update_quv^alpha})(\ref{eq_update_quv^kappa})(\ref{eq_update_quv^omega}).
Note that
the topological network structure is embedded in ${\cal Q}$
whose block-diagonal sizes are allocated by $|\Omega_{u}|-1$
($= d_{u}$ or $d_{u}-1$:
number of links at $u$ in the case of the minimum FVS or VC)
of the submatrix ${\cal Q}_{(e)}$
for links $e:u \rightarrow v$, $u = 1, 2, \ldots, N$,
and its connecting neighbor nodes $v \in \partial u$.

By substituting Eq.(\ref{eq_Qy0}) from Eq.(\ref{eq_Qyinf}),
we have 
\begin{equation}
  {\cal Q} \bm{\Delta y} = \bm{b}(\infty) - \bm{b}(0)
  \stackrel{\rm def}{=} \bm{\Delta b},
  \label{eq_sys_lin_eq}
\end{equation}
where $\varepsilon$-th element of
$\bm{\Delta y} \stackrel{\rm def}{=} \bm{y}(\infty) - \bm{y}(0)$
is $\Delta y_{\varepsilon} = \log r_{e}^{\kappa_{u}}$
from introducing the change rate 
$r_{e}^{\kappa_{u}} \stackrel{\rm def}{=}
q_{e}^{\kappa_{u}}(\infty) / q_{e}^{\kappa_{u}}(1)$ 
and 
\[
\log q_{e}^{\kappa_{u}}(\infty) = \log q_{e}^{\kappa_{u}}(1)
+ \log r_{e}^{\kappa_{u}}.
\]
Remember the definition of $\bm{y}(t)$ by Eq.(\ref{eq_def_y}).

When we consider a vector $\bm{n^{(q)}}$ in the null space
$\{ \bm{n} | {\cal Q} \bm{n} = \bm{0}, \bm{n} \geq \bm{0} \}$
of ${\cal Q}$,
$\bm{\Delta y} + \bm{n^{(q)}}$
is also the solution of Eq.(\ref{eq_sys_lin_eq})
because of
${\cal Q} (\bm{\Delta y} + \bm{n^{(q)}}) = \bm{\Delta b}$.
Since there is only one pair of $\pm 1$ elements in each
row of ${\cal Q}$, $\bm{n^{(q)}}$ is uniquely determined as
$\bm{n^{(q)}} =\{ c_{e_{1}}, \ldots, c_{e_{1}}, \ldots, 
c_{e}, \ldots, c_{e}, \ldots,
c_{e_{2M}}, \ldots, c_{e_{2M}} \}$,
whose elements are divided by $2M$ blocks with any constants
$c_{e} \geq 0$.
In other words,
through multiplying $Q_{(e)}$ of Eq.(\ref{eq_def_Qe}),
\[
(\log q_{e}^{\kappa_{u}} + c_{e}) - (\log q_{e}^{\omega_{u}} + c_{e})
= \log \left(
\frac{q_{e}^{\kappa_{u}} \times e^{c_{e}}}{q_{e}^{\omega_{u}} \times e^{c_{e}}}
\right)
= \log \left( \frac{q_{e}^{\kappa_{u}}}{q_{e}^{\omega_{u}}} \right),
\]
means that
the adding of $\bm{n^{(q)}}$ to $\bm{\Delta y}$
corresponds to any scalar multiples.
However, they disappear by the above division to eliminate
$z_{u \rightarrow v}$
for each link $e: u \rightarrow v$.

In taking into account
stochastic variations of 
initial values $\{ q_{e}^{\kappa_{u}}(0) \}$
generated uniformly at random in the interval $(0, 1)$ 
for $e \in \{ e_{1}, \ldots, e_{2M} \}$ and
$\kappa_{u} \in \Omega_{u}$, $u \in \{ 1, 2, \ldots, N \}$,
we discuss how high is the change rate 
for an equilibrium solution of MP Eqs.
(\ref{eq_update_quv^alpha})(\ref{eq_update_quv^kappa})(\ref{eq_update_quv^omega}).
Essentially, 
each element $q_{e}^{\kappa_{u}}(t)$ is a probability variable
in $(0, 1)$ at any time $t \geq 0$,
the amount of $|q_{e}^{\kappa_{u}}(\infty) - q_{e}^{\kappa_{u}}(0) |$
is small at most $1$.
For the random initial values,
$|| \bm{\Delta b} ||_{2}$ is averagely bounded as $O(1)$,
since the variance of logarithm of finite random variable
becomes a constant (see Appendix A.2).
Remember that
$\bm{\Delta b}$ is defined by
Eqs.(\ref{eq_def_b}) and (\ref{eq_sys_lin_eq}).
When $\tilde{\epsilon} \leq q_{e}^{\kappa_{u}} < 1$ 
is assumed for each link $e:u \rightarrow v$
and state $\kappa_{u} \in \Omega_{u}$,
$\frac{|S_{\kappa_{u}}| \times \tilde{\epsilon}}{|S_{\omega_{u}}|} \leq
\frac{\sum_{\delta'_{u} \in S_{\kappa_{u}}}
    q_{w' \rightarrow u}^{\delta'_{u}}(0)}{\sum_{\delta''_{u} \in S_{\omega_{u}}}
    q_{w' \rightarrow u}^{\delta''_{u}}(0)}
< \frac{|S_{\kappa_{u}}|}{|S_{\omega_{u}}| \times \tilde{\epsilon}}$
is obtained
in the right-hand side of Eq.(\ref{eq_def_b}).
Thus, $b_{\varepsilon}(0)$ have a finite variance,
while $b_{\varepsilon}(\infty)$ is a constant on the assumption
of an equilibrium solution.

Moreover, it is known that
the generalized inverse matrix ${\cal Q}^{\dag}$
gives a solution of Eq.(\ref{eq_sys_lin_eq})
with the minimum $L_{2}$-norm $|| \bm{\Delta y} ||_{2}$
in many solutions
for the underconstraint based on a landscape
$K \times (K+2M)$ matrix  ${\cal Q}$,
\begin{equation}
\bm{\Delta y} = {\cal Q}^{\dag} \bm{\Delta b}
= {\cal Q}^{T} ({\cal Q} {\cal Q}^{T})^{-1} \bm{\Delta b},
\label{eq_yQQTb}
\end{equation}
where ${\cal Q} {\cal Q}^{T}$ becomes
a $K \times K$ block-diagonal matrix, whose each block is
$(|\Omega_{u}| -1) \times (|\Omega_{u}| -1)$ submatrix as follows
\begin{equation}
\begin{pmatrix}
  2      & 1      & \ldots & 1 \\
  1      & 2      & \ldots & 1 \\
  \vdots & \vdots & \ddots & \vdots \\
  1      & 1      & \ldots & 2
\end{pmatrix}.
\label{eq_inv_QQT}
\end{equation}
In general, for a block-diagonal matrix,
the inverse matrix is obtained as
\[
\begin{pmatrix}
  \boxed{B_{1}} &                 &               & &  \\
               & \ddots          &               & \text{\huge{0}} & \\
               &                 & \boxed{B_{i}}  & &  \\
               & \text{\huge{0}} &               & \ddots & \\
               &                 &               & & \boxed{B_{2M}}
\end{pmatrix}^{-1} =
\begin{pmatrix}
  \boxed{B_{1}^{-1}} &                 &               & &  \\
               & \ddots          &               & \text{\huge{0}} & \\
               &                 & \boxed{B_{i}^{-1}}  & &  \\
               & \text{\huge{0}} &               & \ddots & \\
               &                 &               & & \boxed{B_{2M}^{-1}}
\end{pmatrix},
\]
$B_{i}^{-1}$ denotes the inverse of $B_{i}$
for $1 \leq i \leq 2M$.
In considering the order of
$({\cal Q} {\cal Q}^{T})^{-1}$ in Eq.(\ref{eq_yQQTb}),
as $k \stackrel{\rm def}{=} |\Omega_{u}| -1$, 
the inverse of $k \times k$ submatrix of Eq.(\ref{eq_inv_QQT})
is given by 
\[
\begin{pmatrix}
  2      & 1      & \ldots & 1 \\
  1      & 2      & \ldots & 1 \\
  \vdots & \vdots & \ddots & \vdots \\
  1      & 1      & \ldots & 2
\end{pmatrix}^{-1} = \frac{1}{k+1}
\begin{pmatrix}
  k       & -1     & \ldots & -1 \\
  -1      & k      & \ldots & -1 \\
  \vdots  & \vdots & \ddots & \vdots \\
  -1      & -1     & \ldots & k
\end{pmatrix}.
\]

From Eq.(\ref{eq_yQQTb}) and the above discussion,
$\bm{\Delta y}$ is of order
$\frac{1}{\min \{ |\Omega_{u}| \} }$ at most
even in the logarithmic space whose element is
$\Delta y_{\varepsilon} \stackrel{\rm def}{=}
\log (q_{e}^{\kappa_{u}}(\infty) / q_{e}^{\kappa_{u}}(1))$,
because the submatrix of Eq.(\ref{eq_inv_QQT}) is of order
$\frac{1}{k + 1} = \frac{1}{|\Omega_{u}|}$.
According to $|\Omega_{u}| = d_{u} + 1$ or $d_{u}$ for the minimum FVS or VC,
the convergence of state probability may be faster on link
$e:u \rightarrow v$ emanated from node $u$ with
as higher degree $d_{u}$,
although it is not determined by only the probabilities on link
$e: u \rightarrow v$
but depends on ones
(especially at the times $0$ and $\infty$)
on adjacent links $w' \rightarrow u$
with the complex cooperative or competitive 
interactions embedded in $\bm{\Delta b}$.

Thus, 
a solution $\{ q_{e}^{\kappa_{u}}(\infty) \}$ exists in a neighborhood
of any random initial $\{ q_{e}^{\kappa_{u}}(0) \}$
with high probability.
Table \ref{table_corresponding} show the correspondence
in linear theories for our 
MP in product-sum form and learning of neural networks.
Particularly,
the following difference is remarkable.
Once a network is given,
the matrix ${\cal Q}$ is fixed in the case of MP
from initial values chosen uniformly at random.
However even if a neural network is given topologically,
the matrix $X$ is variational
because of the connection weights chosen from a Gaussian
distribution in the case of learning of neural networks
\cite{Amari20a}.

\section{Conclusion}
We study the fast concergence by MP generalized in product-sum
form for finding an approximate solution of combinatorial 
optimization problems such as the minimum FVS \cite{Zhou13}
or VC \cite{Weigt06}.
Actually, the numerical results show
the very fast convergence by MP until only several iterations
less than around ten even for large networks with thousands
nodes and links.
The key contribution is to generalize the MP equations into
a unified product-sum form.
We emphasize that 
the MP is different from BP \cite{Yedidia01}
in sum-product or max-sum form \cite{Shah13}
on a graphical model,
rather its mathmatical framework is related to that in
learning of nueral networks \cite{Amari20a}.
As similar but slightly different way to learning of nueral networks,
a linear theory is applied, and it is derived as a reason of
fast converegence that the equilibrium solution of MP exist
in a neighborhood of initial values in the logorithmic space.
In addition, the effect of degree distribution on the convergence
may be important from the fact that the logarithm of change rate 
$\bm{\Delta y}$ is order $\frac{1}{\min \{ |\Omega_{u}| \}}$,
especially 
$\Omega_{u} = d_{u}$ or $d_{u} - 1$ for the minimum FVS or VC.

To more deeply understand the mechanism,
there still remain several issues as follows.
Even belonging in a same form of product-sum, MP Eqs.
(\ref{eq_update_quv^0})(\ref{eq_update_quv^u})(\ref{eq_update_quv^w})
and 
(\ref{eq_update_puv^0})(\ref{eq_update_puv^*})
are not completely same, and produce slightly different behavior
in Figs. \ref{fig_simTS_FVS} and \ref{fig_simTS_VC}
or in Figs. \ref{fig_AveFreq_FVS} and \ref{fig_AveFreq_VC}.
Also,
varieties of topological network structure seem to affect them
as shown by color lines in these Figures for SF networks of
even similar power-law degree distributions 
such as examples in Table \ref{table_data_source}
but with different $N$, $M$, and $D$.
Since there exist uncountably many network structures
away from SF networks,
it will requires further studies to discover 
the reason of differences.
As the first step,
for a fixed network structure
e.g. randomized networks under a degree distribution
by eliminating other topological properties,
it may be useful to discuss relations between the convergent
behavior and typical sum-forms or number of states with penalty
in classifying what types of product-sum forms can be considered.

On the other hand, it will be expected that 
our discussion is applied to other MP equations
for such as the minimum dominating set \cite{Sun19}
or community detection \cite{Newman23}.
Instead of the cluster variation method \cite{Yedidia01}
for a loopy network,
the extended development of MP by considering
primitive cycles \cite{Newman23} may be useful even with
complex calculations 
to treat the independence more accurately for finding
an unique solution.
However it is out from our approach, or 
we consider the existence of many solutions positively,
because they are feasible solutions near the
optimal as proper approximations.
In addition,
other development of elegant algorithms may be possible 
from information geometric perspective of MP in
product-sum form (see Appendix A.3).

\begin{table}[ht!]
  \caption{Correspondence in linear theories for
    MP in product-sum form and learning of neural network}
  \label{table_corresponding}
  \begin{tabular}{c|c}
    MP in product-sum form & Learnig of neural network \\ \hline
    $K+2M > K$ & $p > n$ \\
    $\log \tilde{\epsilon} \leq y(0)_{\varepsilon} \stackrel{\rm def}{=}
    \log q_{e}^{\kappa_{u}}(1) < 0$, 
    \; $\varepsilon = 1, \ldots, K+2M$
    & $-\infty < v(0)_{i} < \infty, \; i=1, \ldots, p$ \\
    logarithmic change rate vector $\bm{\Delta y}$
    & difference vector $\bm{\Delta v}$ \\ 
    interactions with adjacent links 
    $\bm{\Delta b} \stackrel{\rm def}{=} \bm{b}(\infty) - \bm{b}(0)$
    & error $\bm{e} \stackrel{\rm def}{=} \bm{f}^{*} - X \bm{v}(0)$ \\
    $b_{\varepsilon}(0)$ defined by Eq.(\ref{eq_def_b})
    & $v_{i}(0)$ generated from a Gausian distribution \\   
    landscape $K \times (K+2M)$ block-diagonal matrix ${\cal Q}$
    & landscape $n \times p$ matrix $X$ \\
    with submatrix ${\cal Q}_{(e)}$ 
    & with row vector $\bm{X_{s}}$ of sample input \\
    $q_{e}^{\kappa_{u}}(0)$ is chosen uniformly at random 
    & $X_{si} = \varphi(\bm{w}_{i} \cdot \bm{x}_{s})$ is an iid variable \\  
    $\bm{\Delta b} = {\cal Q} \bm{\Delta y}$
    & $\bm{e} = X \bm{\Delta v}$ \\
    $\bm{\Delta y} = {\cal Q}^{\dag} \bm{\Delta b}$
    & $\bm{\Delta v} = X^{\dag} \bm{e}$ \\
    $\{ \bm{n^{(q)}} | {\cal Q} \bm{n^{(q)}} = \bm{0} \}$
    & $\{ \bm{n^{(x)}} | X \bm{n^{(x)}} = \bm{0} \}$ \\ \hline
  \end{tabular}
\end{table}

\section*{Acknowledgments}
This research was supported in part by 
JSPS KAKENHI Grant Number JP.21H03425.
The author expresses appreciation to Atsushi Tanaka,
and Fuxuan Liao, Jaeho Kim
for discussing the theoretical contents 
and helping the simulations, respectively.

\appendix
\subsection*{A.1 Linear theory for learning of neural networks}
As a citation, 
we briefly explain the linear theory for a neural network
with one hidden layer \cite{Amari20a} to understand similarity
and difference to our discussion.
The scalar output is given by
\[
f(\bm{x}; \bm{\theta}) \stackrel{\rm def}{=}
\sum_{i=1}^{p} v_{i} \varphi(\bm{w}_{i} \cdot \bm{x}),
\]
where $\varphi(z)$ is a bunded activation function,
$\bm{w}_{i} \cdot \bm{x}$ is the inner product of
input $\bm{x}$ and fixed weight $\bm{w}_{i}$ as
$d$-dimensional vectors in $\mathbb{R}^{d}$,
and 
$\bm{v} = (v_{1}, \ldots, v_{i}, \ldots, v_{p})^{T}$
is a $p$-dimensional variable vector in $\mathbb{R}^{p}$
learned as weight parameters between hidden and output layers.
To simplify the discussion, 
each element of $\bm{w}_{i}$ between input and hidden layers 
is fixed and set by a random Gaussian distribuion
in the interval $( -\infty, +\infty )$
with a finite variance $\sigma_{w}^{2} / p$.

By considering a sample set of $n$ inputs alltogether,
the input-output relation is represented by the follwing
systems of linear equatios.
\[
\bm{f} = X \bm{v},
\]
where we assume $n < p$, 
$X$ is a landscape $n \times p$ matrix,
whose element is
\[
X_{si} = \varphi(\bm{w_{i}} \cdot \bm{x_{s}}),
\;\;\; s = 1, \ldots, n; \; i = 1, \ldots, p.
\]
Inputs $\bm{x_{1}}, \ldots, \bm{x_{n}}$
in the training data are randomly
and independently generated
with bounded $| x_{si}|$ for each element.
Therefore, $X$ has stochastic variations.

Since the optimal parameters $\bm{v}^{*}$ have to satisfy
$X \bm{v}^{*} = \bm{f}^{*}$ given as the teacher signal vector,
we have
\[
\bm{f}^{*} = X (\bm{v}(0) + \bm{\Delta v}),
\]
where
$\bm{\Delta v} \stackrel{\rm def}{=} \bm{v}^{*} -\bm{v}(0)$,
and $\bm{v}(0)$ denotes any initial random vector.
For the error vector
$\bm{e} \stackrel{\rm def}{=} \bm{f}^{*} - X \bm{v}(0)$,
the above equation is rewritten as 
\begin{equation}
  \bm{e} = X \bm{\Delta v},
  \label{eq_X_delta-v}
\end{equation}

By using the generalized inverse matrix
$X^{\dag} \stackrel{\rm def}{=} X^{T} (X X^{T})^{-1}$
of $X$, we obtain
\[
\bm{\Delta v} = X^{\dag} \bm{e}.
\]
Note that, in general for fewer constraints than variables,
the exising of some solutions is possible,
and that $X^{\dag}$ gives one of them as the minimum $L_{2}$-norm 
$|| e ||_{2}$ for $|| X X^{\dag} \bm{e} - \bm{e} ||_{2} = 0$.

The minimum norm solution of Eq.(\ref{eq_X_delta-v}) is written as
\begin{equation}
  \bm{\Delta v} = X^{T} (X X^{T})^{-1} \bm{e},
  \label{eq_Delta_v}
\end{equation}
and the generalized solutions are given by
$\bm{\Delta v} + \bm{n^{(x)}}$,
where $\bm{n^{(x)}}$ is an arbitrary null vector belonging to
the null space $\{ \bm{n^{(x)}} | X \bm{n^{(x)}} = \bm{0} \}$.

Moreover,
since the elements of $X X^{T}$ are sum of $p$ iid variables,
the inverse $(X X^{T})^{-1}$ is of order $1/p$.
From Eq.(\ref{eq_Delta_v}) and 
$|| \bm{v_{0}} ||_{2} = \sigma_{v}^{2} = O(1)$, we have
\[
|| \bm{\Delta v} ||_{2} = O\left( \frac{1}{\sqrt{p}} \right).
\]
Thus, a solution $\bm{v_{0}} + \bm{\Delta v}$ exists in a
$(1/\sqrt{p})$-neighborhood of any random initial vector $\bm{v_{0}}$
with high probability.
Such discussion is extended to learning of multilayer neural networks
with variable weight parameters between layers \cite{Amari20a}.

\subsection*{A.2 Bounded variance of logarithmic function}
We show that, for a random variable $x$,
the mean and variance of its logarithmic function are bounded.
We set $0 < \tilde{\epsilon} \leq x \leq X_{max} \approx 1/\tilde{\epsilon}$
and $\tilde{\epsilon} \ll 1$.
At least, in computer simulation with the calculation of $\log x$,
$0 < \tilde{\epsilon} \leq x$ is necessary.
For example, $\log \tilde{\epsilon} \approx - 700$
when $\tilde{\epsilon} \approx 10^{-320}$ in IEEE754 
double-precision floating-point number is applied.

The mean is defined as
\begin{eqnarray}
  \mu_{q} & = & \frac{1}{X_{max} - \tilde{\epsilon}}
  \int_{\tilde{\epsilon}}^{X_{max}} \log x dx \nonumber\\
  & = & \frac{1}{X_{max} - \tilde{\epsilon}}
  \left[ x \log x - x \right]_{\tilde{\epsilon}}^{X_{max}}
  \approx \frac{X_{max}(\log (X_{max}) - 1)}{X_{max} - \tilde{\epsilon}}, \nonumber
\end{eqnarray}
where we assume that the distribution of $x$ is uniformly at random.

Similarly, the variance is defined as
\begin{eqnarray}
  \sigma_{q}^{2} & = &
  \frac{1}{X_{max} - \tilde{\epsilon}} \int_{\tilde{\epsilon}}^{X_{max}}
  \left( \log x - \mu_{q} \right)^{2} dx \nonumber\\
  & = & \frac{1}{X_{max} - \tilde{\epsilon}} \int_{\tilde{\epsilon}}^{X_{max}} (\log x)^{2} dx
  - \frac{2 \mu_{q}}{X_{max} - \tilde{\epsilon}} \int_{\tilde{\epsilon}}^{X_{max}} \log x dx
  + \frac{1}{X_{max} - \tilde{\epsilon}} \int_{\tilde{\epsilon}}^{X_{max}} \mu_{q}^{2} dx. \nonumber
\end{eqnarray}
The third and second terms of above right-hand side are
\[
\frac{X_{max} - \tilde{\epsilon}}{X_{max} - \tilde{\epsilon}}\mu_{q}^{2} -2\mu_{q}
\times \mu_{q} = - \mu_{q}^{2}.
\]
For a permutation integral, we set $z = \log x$.
Then, the first term is
\begin{eqnarray}
  \frac{1}{X_{max} - \tilde{\epsilon}} \int_{\log \tilde{\epsilon}}^{\log X_{max}} z^{2} e^{z} dz 
  & = & \frac{1}{X_{max} - \tilde{\epsilon}} \left[ (z^{2} -2 z + 2) e^{z}
    \right]_{\log \tilde{\epsilon}}^{\log X_{max}} \nonumber\\
  & \approx &
  \frac{1}{X_{max} - \tilde{\epsilon}} (\log X_{max})^{2} X_{max}. \nonumber
\end{eqnarray}
Therefore, we obtain the bounded constant value 
\[
\sigma_{q}^{2} \approx
\frac{1}{X_{max} - \tilde{\epsilon}} (\log X_{max})^{2} X_{max}
- \mu_{q}^{2}.
\]

\subsection*{A.3 Information geometric perspective}
In the $(m-1)$-dimensional statistical manifold over
the finite discrete set $\chi = \{ 1, 2, \ldots, x, \ldots, m \}$, 
we consider a $n$-dimensional submanifold called
exponential family
$S_{E} = \{ p_{E}(x; \theta) | x \in \chi, \theta^{i} \in \mathbb{R} \}$
with parameter
$\bm{\theta} = (\theta^{1}, \ldots, \theta^{i}, \ldots, \theta^{n})$,
$n < m-1$.
The probability distribution is represented in the following
normal form \cite{Amari00}.
\[
p_{E}(x; \theta) \stackrel{\rm def}{=} \exp \left\{ C(x) +
\sum_{i=1}^{n} F_{i}(x) \theta^{i} - \psi(\theta) \right\},
\]
\[
\psi(\theta) \stackrel{\rm def}{=} \log \left\{ \sum_{x \in \chi} \exp
  \left( C(x) + \sum_{i=1}^{n} F_{i}(x) \theta^{i} \right) \right\}.
  \]
  
Without loss of generality,
we chose $C(x) = \log p_{0}(x) = 0$,
$F_{i}(x) = \log p_{i}(x) - \log p_{0}(x) = \log p_{i}(x)$,
where $p_{0}(x)$ is the uniform distribution,
and $p_{i}(x) > 0$ is a function on $x \in \chi$ 
for each $i = 1, 2, \ldots, n$.
Then, $p_{E}(x; \theta)$ is rewritten as
\[
p_{E}(x; \theta) = \frac{\Pi_{i=1}^{n}
  p_{i}(x)^{\theta^{i}}}{\sum_{x \in \chi}
  \Pi_{i=1}^{n} p_{i}(x)^{\theta^{i}}}.
\]
Moreover, after easy calculations with logarithmic transformation,
we obtain the system of linear equations \cite{Hayashi98}
\[
\sum_{i=1}^{n} \left( F_{i}(x) - F_{i}(m) \right) \theta^{i}
= \log \left( \frac{p_{E}(x; \theta)}{p_{E}(m; \theta)} \right),
\]
\[
F_{i}(x) - F_{i}(m) = \log \left( \frac{p_{i}(x)}{p_{i}(m)} \right).
\]

For each link $u \rightarrow v$,
$q_{u \rightarrow v}^{\kappa_{u}}$ in Eq.(\ref{eq_update_quv^kappa})
is corresponded to $p_{E}(x; \theta)$
in the mapping of $n$ and $k_{u}-1 = | \partial u \backslash v |$, 
$\chi = \{ 1, \ldots, x, \ldots, m \}$ and
$\Omega_{u} = \{ \alpha_{u}, \ldots, \kappa_{u}, \ldots, \omega_{u} \}$,
$F_{i}(x)$ and the logarithm of the numerator
in the right-hand side of Eq.(\ref{eq_update_quv^alpha})
(\ref{eq_update_quv^kappa}) or (\ref{eq_update_quv^omega}),
with $\bm{\theta} = (1, 1, \ldots, 1)$.
In other words,
the basis function $F_{i}(x)$ is arranged 
according to the updating by MP 
Eq.(\ref{eq_update_quv^alpha})
(\ref{eq_update_quv^kappa}) or (\ref{eq_update_quv^omega})
which depends on the state prababilities
on other adjecent links
$w \rightarrow u$, $w \in \partial u \backslash v$.

Thus, from the above explanation,
we can regard $\{ q_{u \rightarrow v}^{\kappa_{u}}(t) \}$
for each link $e: u \rightarrow v$ 
as an exponential family.
However,
it is different from the information geometric explanation
for the sum-product or max-sum form \cite{Shah13}
of MP called BP
applied to a graphical model \cite{Yedidia01},
in which 
the parameter $\bm{\theta}$ is arranged 
according to the updating of MP
\cite{Ikeda04}.

On the other hand,
another example of exponential family is Boltzman machine
\cite{Amari00, Byre92} as one of the well-known stochastic
neural networks.
Moreover,
such as EM, independent component analysis,
and natural gradient method, 
elegant algorithms have been provided from information
geometric foundations \cite{Amari95}.

\end{document}